\documentclass[sn-nature,Numbered]{sn-jnl}

\usepackage{graphicx}%
\usepackage{multirow}%
\usepackage{amsmath,amssymb,amsfonts}%
\usepackage{amsthm}%
\usepackage{mathrsfs}%
\usepackage[title]{appendix}%
\usepackage{xcolor}%
\usepackage{textcomp}%
\usepackage{manyfoot}%
\usepackage{booktabs}%
\usepackage{algorithm}%
\usepackage{algorithmicx}%
\usepackage{algpseudocode}%
\usepackage{listings}%

\usepackage{multibib}
\newcites{supp}{Methods References}
\usepackage{caption}

\theoremstyle{thmstyleone}%
\theoremstyle{thmstyletwo}%

\theoremstyle{thmstylethree}%

 \usepackage{aas_macros}

\def\lesssim{\mathrel{\hbox{\rlap{\hbox{\lower4pt\hbox{$\sim$}}}\hbox{$<$}}}}
\def\gtrsim{\mathrel{\hbox{\rlap{\hbox{\lower4pt\hbox{$\sim$}}}\hbox{$>$}}}}

\def\micron{\hbox{$\mu$m}}

\definecolor{ao}{rgb}{0.0, 0.5, 0.0}

\newcommand\reduline{\bgroup\markoverwith{\textcolor{red}{\rule[-0.5ex]{2pt}{0.4pt}}}\ULon}

\newcommand{\Msun}{\hbox{M$_\odot$}}

\newcommand{\OVI}{[\hbox{{\rm O}\kern 0.1em{\sc vi}}]}

\newcommand{\NV}{\hbox{{\rm N}\kern 0.1em{\sc v}}}
\newcommand{\SiIV}{\hbox{{\rm Si}\kern 0.1em{\sc iv}}}
\newcommand{\OIV}{[\hbox{{\rm O}\kern 0.1em{\sc iv}}]}
\newcommand{\NIV}{[\hbox{{\rm N}\kern 0.1em{\sc iv}}]}
\newcommand{\CIV}{\hbox{{\rm C}\kern 0.1em{\sc iv}}}
\newcommand{\HeII}{\hbox{{\rm He}\kern 0.1em{\sc ii}\kern 0.1em{$\lambda1640$} }}
\newcommand{\OIII}{[\hbox{{\rm O}\kern 0.1em{\sc iii}}]{$\lambda5007$}}
\newcommand{\OIIId}{[\hbox{{\rm O}\kern 0.1em{\sc iii}}]{$\lambda4959\lambda5007$}}

\newcommand{\NIII}{[\hbox{{\rm N}\kern 0.1em{\sc iii}}]}
\newcommand{\AlIII}{\hbox{{\rm Al}\kern 0.1em{\sc iii}}}
\newcommand{\SiIII}{\hbox{{\rm Si}\kern 0.1em{\sc iii}}}
\newcommand{\CIII}{\hbox{{\rm C}\kern 0.1em{\sc iii}]}}
\newcommand{\NeIV}{[\hbox{{\rm Ne}\kern 0.1em{\sc iv}}]}
\newcommand{\MgII}{\hbox{{\rm Mg}\kern 0.1em{\sc ii}}}

\newcommand{\CII}{[\hbox{{\rm C}\kern 0.1em{\sc ii}]}}

\newcommand{\He}{\hbox{{\rm He}\kern 0.1em{\sc ii}\kern 0.1em{$\lambda1640\lambda4686$}}}

\newcommand{\SII}{[\hbox{{\rm S}\kern 0.1em{\sc ii}}]$\lambda6717\lambda6731$}
\newcommand{\NII}{[\hbox{{\rm N}\kern 0.1em{\sc ii}}]}
\newcommand{\OII}{[\hbox{{\rm O}\kern 0.1em{\sc ii}}]}

\newcommand{\MgI}{\hbox{{\rm Mg}\kern 0.1em{\sc i}}}
\newcommand{\FeII}{\hbox{{\rm Fe}\kern 0.1em{\sc ii}}}

\newcommand{\OI}{\hbox{{\rm O}\kern 0.1em{\sc i}}}
\newcommand{\NeII}{[\hbox{{\rm Ne}\kern 0.1em{\sc ii}}] }
\newcommand{\NaI}{[\hbox{{\rm Na}\kern 0.1em{\sc i}}] }
\newcommand{\NeIII}{[\hbox{{\rm Ne}\kern 0.1em{\sc iii}}] }

\begin{document}

\title[]{A massive galaxy that formed its stars at $z \sim 11$}

\author*[1]{\fnm{Karl} \sur{Glazebrook}}\email{kglazebrook@swin.edu.au}
\author[1]{\fnm{Themiya} \sur{Nanayakkara}}
\author[2]{\fnm{Corentin} \sur{Schreiber}}
\author[3,9,10]{\fnm{Claudia} \sur{Lagos}}
\author[1]{\fnm{Lalitwadee} \sur{Kawinwanichakij}}
\author[1]{\fnm{Colin} \sur{Jacobs}}
\author[1]{\fnm{Harry} \sur{Chittenden}}
\author[3]{\fnm{Gabriel} \sur{Brammer}}
\author[1]{\fnm{Glenn G.} \sur{Kacprzak}}
\author[1]{\fnm{Ivo} \sur{Labbe}}
\author[4]{\fnm{Danilo} \sur{Marchesini}}
\author[5]{\fnm{Z.~Cemile} \sur{Marsan}}
\author[3,6]{\fnm{Pascal A.} \sur{Oesch}}
\author[7]{\fnm{Casey} \sur{Papovich}}
\author[11]{\fnm{Rhea-Silvia} \sur{Remus}}
\author[8,9]{Kim-Vy H. Tran}
\author[1]{\fnm{James} \sur{Esdaile}}
\author[10]{\fnm{Angel} \sur{Chandro-Gomez}}


\affil*[1]{\orgdiv{Centre for Astrophysics and Supercomputing}, \orgname{Swinburne
  University of Technology}, \orgaddress{\street{P.O. Box 218}, \city{Hawthorn}, \postcode{3122}, \state{VIC}, \country{Australia}}}
  
\affil[2]{\orgname{IBEX Innovations},  \orgaddress{\street{Sedgefield}, \city{Stockton-on-Tees}, \postcode{TS21 3FF}, \country{United Kingdom}}}

\affil[3]{\orgdiv{Cosmic DAWN Center, Niels Bohr Institute}, \orgname{University of Copenhagen},  \orgaddress{\street{Jagtvej 128}, \city{Copenhagen N}, \postcode{DK-2200}, \country{Denmark}}}

\affil[4]{\orgdiv{Physics and Astronomy Department}, \orgname{Tufts University}, 
\orgaddress{574 Boston Avenue}, \city{Medford}, \state{MA}, \postcode{02155}, \country{USA}}

\affil[5]{\orgdiv{Department of Physics and Astronomy}, \orgname{York University}, 
\orgaddress{4700 Keele Street}, \city{Toronto}, \state{ON}, \postcode{M3J 1P3}, \country{Canada}}

\affil[6]{\orgdiv{Department of Astronomy}, \orgname{University of Geneva},  \orgaddress{\street{Chemin Pegasi 51}, \city{Versoix}, \postcode{CH-1290}, \country{Switzerland}}}

\affil[7]{\orgdiv{Department of Physics and Astronomy, and George P. and Cynthia Woods Mitchell Institute for Fundamental Physics and Astronomy}, \orgname{Texas A\&M University},  \orgaddress{\city{College Station}, \state{TX} \postcode{77843-4242}, \country{USA}} }

\affil[8]{School of Physics, University of New South Wales, Kensington, Australia}
\affil[9]{ARC Centre for Excellence in All-Sky Astrophysics in 3D}

\affil[10]{\orgdiv{International Centre for Radio Astronomy Research}, \orgname{University of Western Australia}, \orgaddress{\street{7 Fairway}, \city{Crawley}, \postcode{6009}, \state{WA}, \country{Australia}}}

\affil[11]{\orgdiv{Universit\"ats-Sternwarte, Fakult\"at für Physik, Ludwig-Maximilians-Universit\"at M\"unchen},\orgaddress{\street{ Scheinerstr. 1}, \postcode{81679}, \city{M\"unchen}, \country{Germany}}}

\maketitle

\newcommand{\RR}{$^{\bf \color{red}ref\ }$}
\newcommand{\red}[1]{{\color{red}#1}}

\textbf{
The formation of galaxies by gradual hierarchical co-assembly of  baryons and cold dark matter halos is a fundamental paradigm underpinning modern astrophysics\cite{Blum84,Somerville2015}
and predicts a strong decline in the number of massive galaxies at early cosmic times\cite{KG04,Behroozi2018,BK23}. 
Extremely massive quiescent galaxies (stellar masses $>10^{11}$\Msun) have now been observed as early as 1--2 billions years after the Big 
Bang\cite{Marsan+2015,G17,S18,Forrest2019,Saracco+2020,Forrest2020,Valentino2020,Antwi-Danso2023}; these are extremely constraining
on  theoretical models as they form 300--500 Myr earlier and only some models can form massive galaxies this early
\cite{Merlin2019,Valentino2020}. Here we report on the spectroscopic observations
with the James Webb Space Telescope of a massive quiescent galaxy ZF-UDS-7329 at redshift $3.205\pm 0.005$ that eluded deep
ground-based spectrscopy\cite{S18}, is significantly redder than typical and whose
spectrum reveals features typical of much older stellar populations. Detailed modelling shows the stellar population
formed around 1.5 billion years earlier in time ($z \sim 11$) at an epoch when dark matter halos of 
sufficient hosting mass have not yet assembled in the standard scenario\cite{Behroozi2018,BK23}. 
This  observation may point to the presence of  undetected populations of early galaxies and
the possibility of 
significant gaps in our understanding of early stellar populations, galaxy formation and/or the nature of dark matter. 
}

\bigskip

\begin{figure}%
\includegraphics[width=14.5cm]{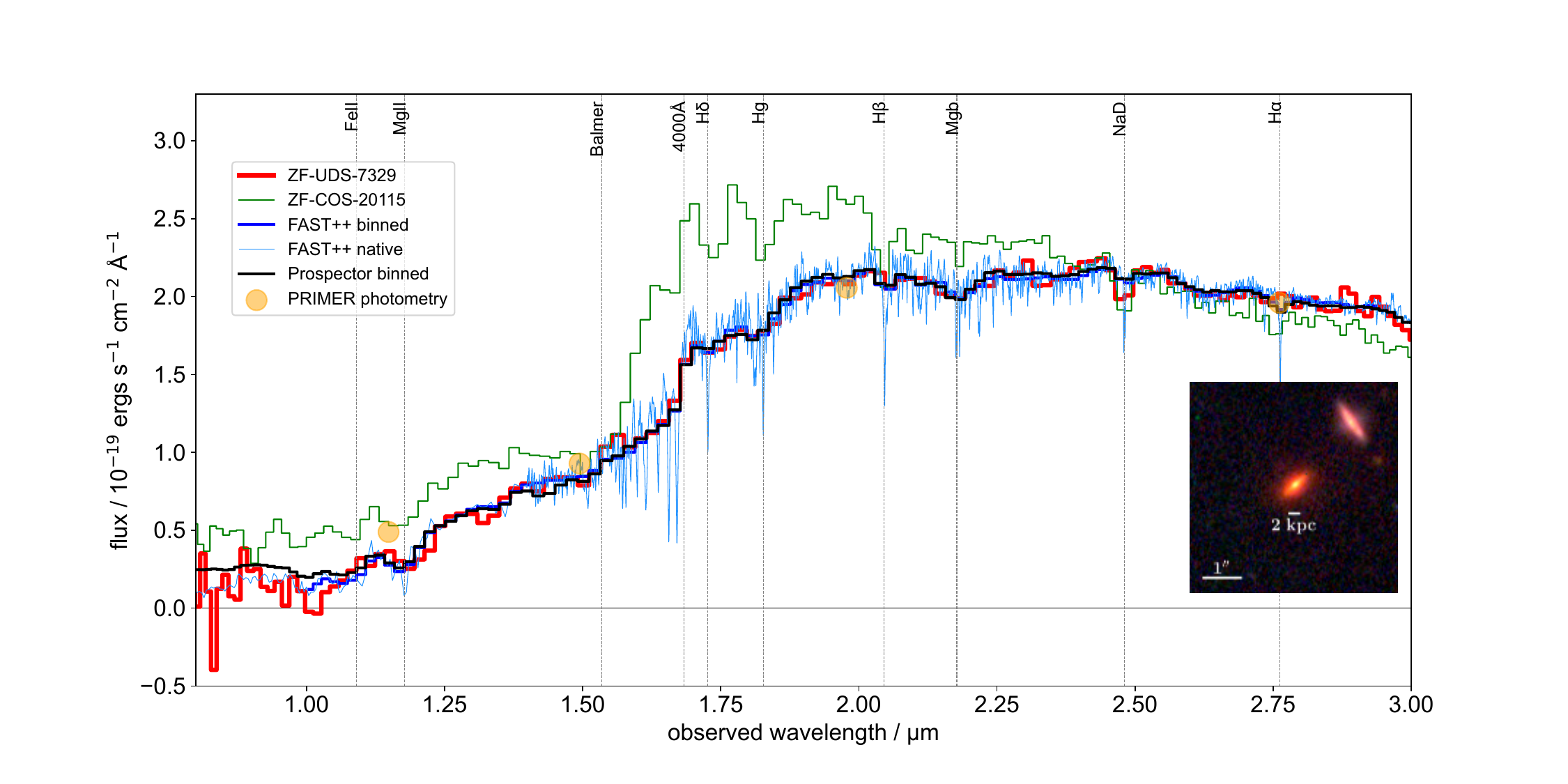}
\caption{The JWST NIRSpec spectrum of galaxy ZF-UDS-7329. This covers rest frame 0.2--0.7\micron\ and is compared to the best fit FAST++ model which is shown binned to the varying pixel size
and resolution of the PRISM spectrum as well as at its original native resolution. Wavelengths of common absorption and emission lines
are marked.  It can be seen that the 4000\AA\ break is well developed and the spectrum is dominated
by old cool stellar features. The Prospector fit is very similar. 
For comparison our recent NIRSpec spectrum of ZF-COS-20115 which has a very similar stellar mass
is shown (corrected to the same redshift), this galaxy has been well studied and is significantly younger\cite{G17,S18} being quenched for $\sim$500 Myr. Its 
spectrum is more typical of post-starburst quiescent galaxies at this epoch showing a bluer break due to the Balmer limit and strong Balmer absorption lines from hot A stars that remain in the spectrum at this younger age.
The PRIMER photometry is also shown and the inset shows the PRIMER image (colour composite of F115W, F200W, and F444W covering rest frame
optical bands) Note the spectral and photometric errors from noise are too small to plot in this figure.
\label{fig:spec}}
\end{figure}

\begin{figure}%
\includegraphics[width=14.5cm]{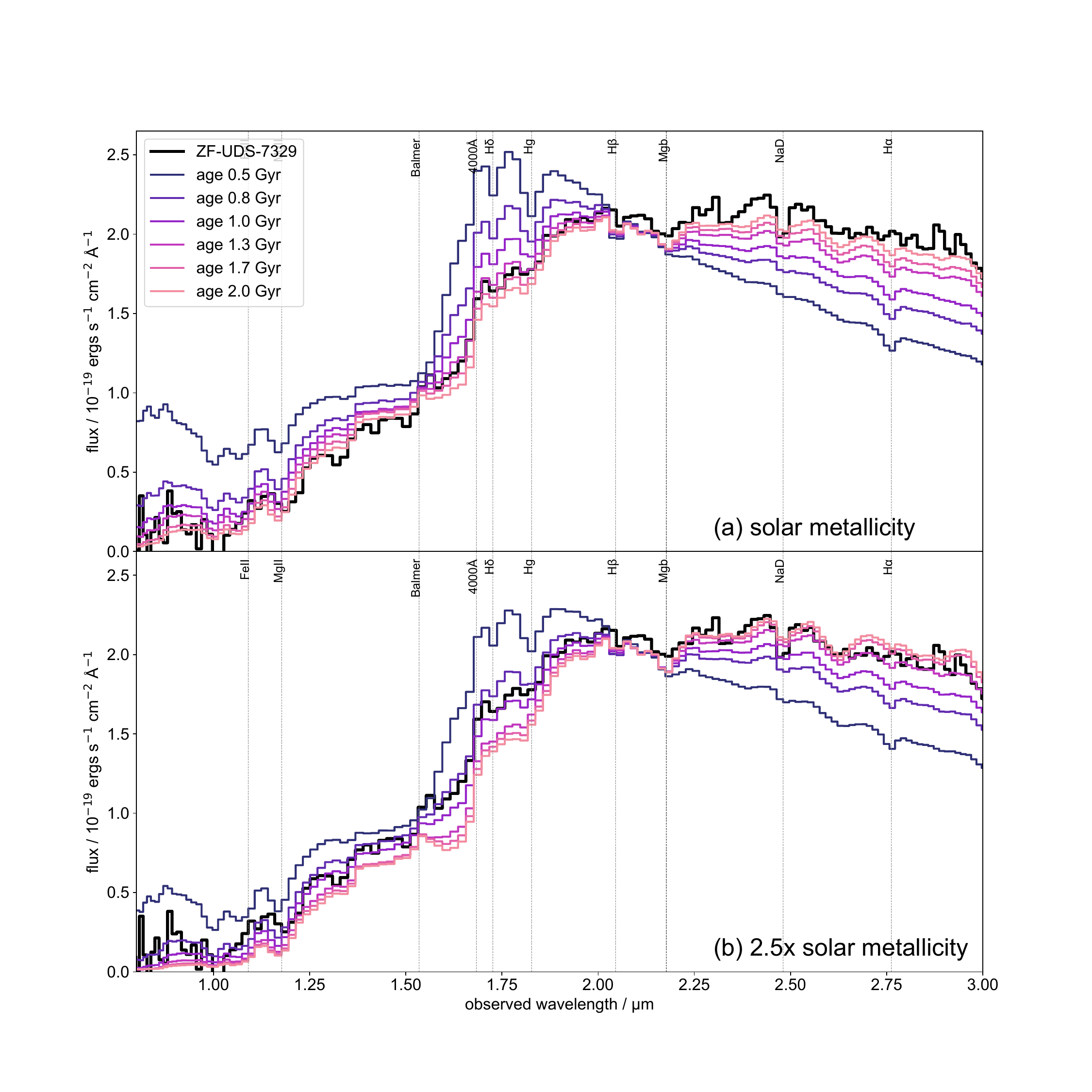}
\caption{The effect of age and metallicity on stellar populations. Simple Stellar Population models \cite{BC03} are compared 
with the 
JWST spectrum of ZF-UDS-7329. It can be seen how the transition from a post-starburst type spectrum 
with a Balmer break and Balmer lines from young A stars  to an evolved population with a 4000\AA\ break happens at $>800$ Myr.  Panel (a) shows how the model at $\simeq$ 1.7 Gyr is an excellent match to the data for solar metallicity [Fe/H]=0. It can be seen how the strong Balmer lines disappear. 
Panel (b) shows supersolar metallicity [Fe/H]=0.05. As expected due to age--metallcity degeneracy the overall shape best matches at a younger age ($\simeq 1.0$ Gyr); however the detailed features
of the spectrum are not well reproduced particularly around the 4000\AA\ break, this is expected as this feature is produced by numerous metal absorption lines. Overall the rest frame optical spectrum
is not consistent with a supersolar model.
\label{fig:SSPs}}
\end{figure}

The spectrum of  ZF-UDS-7329 taken with the NIRSPEC prism disperser  (spectral resolution of 40--350) is shown in (Figure~\ref{fig:spec}) and shows clear absorption
features such as the 4000\AA\ break and Mg\,b 5174\AA\ typical of old $>1$ Gyr stellar populations. To quantitively measure the age we considered a range of approaches. First we fitted a grid of FAST++ parametric star formation history models\cite{S18} to the spectrum and photometry (see Methods for details) to derive a stellar
mass of $1.24\pm 0.09\times 10^{11}$ \Msun\ (consistent with the photometric estimate from reference\cite{S18}; hereafter S18), stellar metallicity [Fe/H] of $0.021\pm 0.005$,  and an age $t_{50}$ (defined as when
50\% of the stellar mass had formed). We find $t_{50}=1.52\pm 0.16$ Gyr which is significantly older than other quiescent galaxies at this epoch\cite{Marsan+2015,G17,S18,Forrest2019,Saracco+2020,Forrest2020,Valentino2020,Antwi-Danso2023}. This accords with the visual
presence of the 4000\AA\ break, which only develops in stellar populations at ages  $\gtrsim$ 800 Myr as illustrated in Figure~\ref{fig:SSPs}.  The spectrum is clearly
different, even at our relatively low spectral resolution ($R=51$ at the break), from younger coeval post-starbursts which show a broader Balmer 3640\AA\ break and strong Balmer absorption lines. This can be seen by the direct comparison in Figure~\ref{fig:spec} with the well studied\cite{G17,S18} $z=3.717$ post-starburst quiescent galaxy ZF-COS-20115. The fitted dust attenuation is small ($<$0.3 mags in the rest frame $V$-band); we note our sub-mm observations of other objects in the S18 sample
have also found low dust and gas content\cite{Suzuki2022}. NIRCAM JWST images of ZF-UDS-7329 reveal a compact, edge-on red disk
 (effective radius $1.15\pm 0.08$ kpc at rest frame 1\micron) that show no evidence for any younger star-forming clumps. 
 
\begin{figure}%
\includegraphics[width=13.5cm]{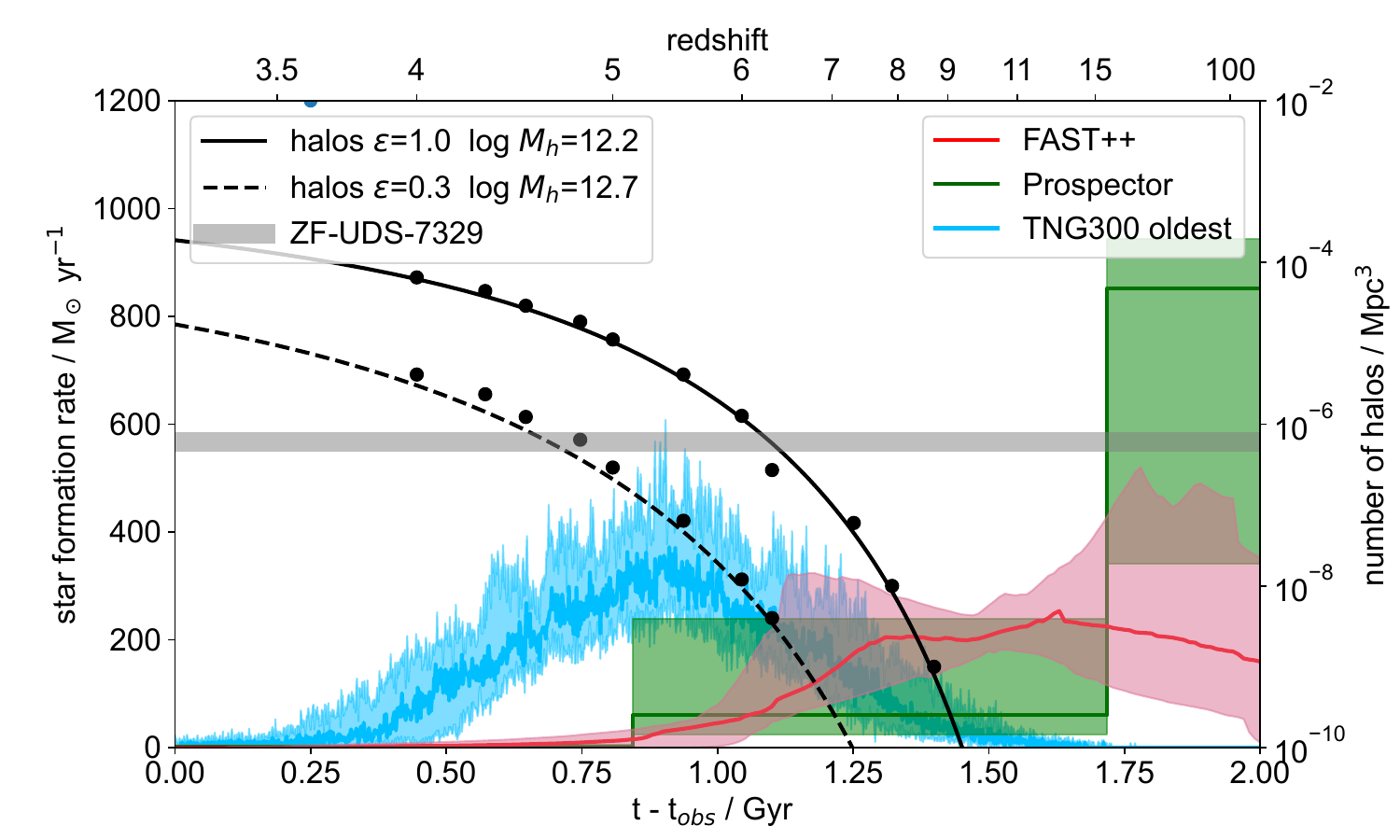}
\caption{Comparison of star formation histories with dark matter halo assembly histories.
The red and green lines represent FAST++ and Prospector star
formation histories inferred from our fitting to ZF-UDS-7329 with the shading showing 16 and 84 percentile ranges in each time bin derived from our Monte Carlo procedure. The FAST++ fits are of
dual exponential functional  form whereas the Prospector ones are stepwise. Both modelling procedures, utilising different spectral
libraries and codes, produce broadly similar star formation histories peaked towards very old ages with the majority of the stars forming at ages $\gtrsim 1.5$ Gyr. 
The blue band shows the median star formation history (with 16/84 percentiles) of the 20 oldest massive quiescent galaxies in TNG300 at $z=3$ for comparison. The evolving 
density of massive cold dark matter halos calculated analytically are shown as solid and dashed black lines for two different halo masses set by two choices of baryon conversion efficiency. The round circles show the
$z>4$ results from the FLAMINGO dark matter only simulation for the same masses. The estimated space density derived from this
galaxy is shown as a horizontal grey band. 
\label{fig:sfh}}
\end{figure}

The old age implies a formation at $z\sim 11$ and large volume simulations show that single galaxies of this mass have not yet formed at this time\cite{Kannan_2023}. To understand the potential
systematic errors we also fit the spectrum using the code Prospector\cite{Johnson2021a}  that has an alternative set of stellar tracks and spectral libraries. This allows us to explore
non-parametric star-formation histories which is  important as it is possible for 
parametric histories to misestimate the mass-weighted ages of galaxies\cite{Carnall2019a,Leja2020a,Seuss2022}. Prospector gives a consistent stellar mass,
metallicity and low attenuation.
We show the reconstruction of the star formation history (SFH) from both codes in Figure~\ref{fig:sfh} with associated uncertainties inferred by Monte Carlo quantiles in each time bin. 
Both are consistent with old ages with the galaxy forming over 200--400 Myr  and all star formation ceasing 1 Gyr before the epoch of observation; Prospector prefers even older populations (up to 1.8 Gyr). Our modelling includes a range of stellar metallicity up to [Fe/H$=$0.05];
these supersolar values are seen in some objects at these redshifts\cite{Saracco+2020} and due to approximate age-metallicity degeneracies could produce younger age solutions. For our spectrum
[Fe/H]$=$0.05 is excluded as a poor fit at 98\% confidence, 
nevertheless if we force this value in our fitting we still need minimum ages $t_{50}\sim 1$ Gyr. 
This fit is able to reproduce the overall red colour of the spectrum; however it differs in the detailed reproductions of the spectral features across the range 1--3\micron. It can be seen in Figure~\ref{fig:SSPs} that
high metallicity populations exhibit a different shape around the 4000\AA\ break due to the increase Fe absorption.

Can such a massive galaxy theoretically be formed at $z\sim 11$? 
The details of galaxy formation within dark matter halos is complex, but a boundary condition can be placed\cite{Behroozi2018} as the most massive halos have a strongly declining abundance with redshift.
Simulations show that haloes
with masses $>10^{11}$ \Msun\ closely track the universal cosmic baryon fraction\cite{Ayromlou2023} of 16\% at $z>2$; this is lowered by feedback but only at late
times and low masses.\cite{Wright2020}. 
We assume this tracking and further assume 100\% conversion of halo baryons in to stars
could be achievable\cite{Behroozi2018} to calculate a minimum halo mass. We also compare with a 30\% conversion 
which is about the maximum baryon-to-stellar conversion efficiency at $3<z<4$ according to semi-empirical estimates of the stellar-halo mass relation\cite{Behroozi+19};
this gives a higher halo mass. For this comparison we need to use the {\it formed\/} stellar mass,
which is higher than the observed residual stellar mass by 0.17--0.22 dex in our fits; we adopt a formed stellar mass of $2.5\times 10^{11}$\Msun\ for this comparison.
First we use an analytic model \cite{Reed2007} to compute the halo mass function as function of redshift;  secondly we directly compute
the halo mass function from the FLAMINGO Gpc$^3$ simulation\cite{schaye2023flamingo}.
We can approximate the space density of galaxies like ZF-UDS-7329 by dividing the quoted space density in S18 by 24 (for 1
 in 24 objects), this gives  $6\times 10^{-7}$ Mpc$^{-3}$. This comparison is shown in Figure~\ref{fig:sfh} where we compare the lookback times
 when the stars form in ZF-UDS-7329 with when halos of suitable mass and number density assemble.  It can be seen that this does not occur until
 $z\sim 6$.  To reconcile this we need to have a significant fraction of the stars form more recently than 1 Gyr; however 
 this is excluded by the old age and star-formation history fits.

We can also compare with the actual stellar mass of objects formed in simulations. In the THESAN reionisation simulations\cite{Kannan_2021} of cubic side length 95.5 comoving Mpc, the stellar mass function at $z=10$ extends to $10^{9}$ \Msun\ at most with a corresponding space density of  $10^{-5}$ Mpc$^{-3}$  such galaxies.
In the MillenniumTNG simulation \cite{Kannan_2023} (740 Mpc side), the highest stellar mass of a galaxy at $z=11$ is just shy of $10^{10}$ \Msun. If direct ancestors of ZF-UDS-7329 existed in these simulations there would be $\sim 200$ in this
volume. 

We next consider other possible
systematic errors and erroneous assumptions. Our spectral fitting procedure for the stellar ages marginalises over a broad
range of star formation histories, including younger burst populations, and the latter are heavily excluded in all viable fits. In ZF-UDS-7329
there is no evidence for any strong emission lines such as might be associated with non-stellar sources
such as Active Galactic Nuclei or residual star-formation. It is significant that even with very high signal:noise ratio (SNR) of 50--70 across the rest-frame optical
we are able to obtain a reduced $\chi^2\sim$ 2 with the fits reproducing 
all the spectral variations. As noted before a  high-metallicity solution [Fe/H]=0.05 is formally a poor fit; even then this brings $t_{50}$ down to 1 Gyr ($z\sim 6$) which 
still requires a 100\% baryon conversion efficiency that has not been seen in any galaxy.
We have assumed a conventional Chabrier Initial Mass Function (IMF); there is some evidence for IMF 
evolution in $z>3$ quiescent galaxies but the effect is at most a factor of two lower in stellar mass\cite{Esdaile+2021,Forrest+2022} which is not enough to 
make a significant difference to our conclusions. 
A more exotic IMF assumption could drastically lower the inferred stellar masses
and there could be other issues such as large mismatches between local templates and early stellar populations.

In $\Lambda$CDM  formation models the prediction is that such a galaxy would have to build up from numerous mergers in its recent past; however this  still predicts young ages
as star formation tracks halo growth.  We quantitively  test the limit of the theoretical expectation
by selecting 283 quiescent galaxies from the Illustris TNG300 simulation\cite{IllustrisTNG} with stellar mass $>10^{10.8}$\Msun\ and specific star formation
rate $<0.15$ Gyr$^{-1}$ (following S18) and tracing their star formation histories through their halo progenitor trees. Additionally we compute $t_{50}$ for each galaxy
and select the 20 oldest ones; this approximates the number density of ZF-UDS-7329 and our pre-selection of the oldest object. 
We show the median star formation history of these 20 objects in Figure~\ref{fig:sfh}; even 
with this extreme assumption we find $t_{50} = 890\pm 50$ Myr with a significant amount of  star formation younger than 1 Gyr; which is a marked contrast to our galaxy. The halo mass
growth follows a similar trajectory. 

Finally we could consider if more exotic cosmological models  can  produce massive halos earlier in cosmic
history. Though there is no obvious alternative to 
vanilla cold dark matter which would clearly do this we note that exotic dark matter models can exert a strong influence on early
galaxy formation formation \cite{Dayal+2015,Maio+Viel+2023}. However these do not necessarily
go in the right direction, for example popular Warm Dark Matter models suppress the abundance of lower mass galaxies but make little
difference at the high-mass end\cite{Lin+2023}. Modifying the initial power spectrum by introducing a `blue tilt' would be one potential solution\cite{Parashari+Laha2023,Padmanabhan_2023}
as is `Early Dark Energy'\cite{BK23} or early seeds from massive primordial black holes\cite{LB22}.

The apparent formation of ZF-UDS-7329 at such an early epoch points to significant problems with our current paradigms of early stellar populations and of galaxy formation. It may even
violate the basic constrain of their being sufficiently massive dark matter halos. We would speculate that some combination of multiple of these effects is likely responsible.
Purely empirically this picture of rapid and early galaxy formation is consistent with the direction of other early results from JWST such as the unexpected overabundance of very luminous $10<z<12$  star-forming galaxies, identified photometrically,  compared to simulations\cite{schaye2023flamingo}. Many of these are now spectroscopically confirmed\cite{Curtis-Lake+2023,Fujimoto+2023} though we note their stellar masses extremely
uncertain and very bursty star-formation may be an alternative explanation of the UV luminosity function\cite{Sun2023}. There is also the photometric identification of six candidate
star forming galaxies with high stellar mass ($\sim 10^{10}$\Msun) \cite{Labbe+2022} at $7<z<9$. This  comparison is particularly interesting as these candidates also stress basic
halo mass constraints\cite{BK23}; however they have yet to be spectroscopically confirmed except for two galaxies which turned out to be significantly lower mass.\cite{Fujimoto+2023} ZF-UDS-7329 is  60\% more massive than the most massive candidate in that sample (and a factor of 9 more massive than the median). A plausible scenario could be constructed in which these are a population of ancestors that merge to make galaxies like ZF-UDS-7329. 

We have presented a galaxy at $z=3.205$ whose very old age, rapid formation and extreme stellar mass are difficult to reconcile with current paradigms of early galaxy formation; these may require significant revision. The main limitation of this work is of course the inferences are based on the discovery of a single object. The inferred sky density is only of order 8 deg$^{-2}$
at $3<z<4$ so they are a rare class of source. The next step is the photometric discovery of more such galaxies to be certain of their abundance, and higher spectral resolution 
observations to measure more constraining elemental abundances and stellar ages. Fortunately both of these are well within the 
capability of JWST\cite{Carnall_NIRSPEC_2023}; if their existence is a challenge to the $\Lambda$CDM galaxy formation paradigm this will be resolved definitively soon.

\bigskip
\bigskip

\backmatter

\bmhead{Acknowledgments}
K.G. thanks Roberto Abraham for assistance in fetching data on $z=0$ comparison galaxies from Canadian archives and Jarle Brinchmann for inspiring discussions on the 0.94\micron\ bump.
This work is based on observations made with the NASA/ESA/CSA James Webb Space Telescope. The data were obtained from the Mikulski Archive for Space Telescopes at the Space Telescope Science Institute, which is operated by the Association of Universities for Research in Astronomy, Inc., under NASA contract NAS 5-03127 for JWST. These observations are associated with program 2565. We thank all the hard work of the JWST team which made this great observatory possible.
We thank Michael Maseda and Allison Strom for helpful discussions during the data reduction process.
T.N., K. G., and C.J. acknowledge support from Australian Research Council Laureate Fellowship FL180100060. 
This work has benefited from funding from the Australian Research Council Centre of
Excellence for All Sky Astrophysics in 3 Dimensions (ASTRO 3D), through project number CE170100013.
C.P. acknowledges generous support from Marsha L. and Ralph F. Schilling and from the George P. And Cynthia Woods Mitchell Institute for Fundamental Physics and Astronomy. 
The Cosmic Dawn Center is funded by the Danish National Research Foundation (DNRF) under grant DNRF140. 
P.O. is supported by the Swiss National Science Foundation through project grant 200020\_207349. This work received funding from the Swiss State Secretariat for Education, Research and Innovation (SERI). 

\bmhead{Author Contributions}
K.G. did all the final data analysis, made the figures and wrote the manuscript. T.N. reduced the NIRSpec data and ran FAST++ and Prospector model fits. 
C.S. originally identified ZF-UDS-7329 as an interesting object  in S18 and analysed deep ground based spectra that failed to secure a redshift but motivated the JWST program. C.S. also added the LSF
functionality to FAST++ and {\tt slinefit} codes for this paper.
C.L. and A.C-G. provided the comparisons with halos in simulations. L.K. processed the NIRCam data. H.C. did TNG300 and THESAN comparisons.
All other authors contributed to the scientific discussions in the proposal and manuscript.

\bmhead{Data Availability}

We make the full JWST spectrum of ZF-UDS-7329 and associated Line Spread Function available in the CSV format small file {\tt final\_spectrum\_pub.csv}. This is the Source Data for Figures~\ref{fig:spec}, \ref{fig:SSPs} and Extended Data Figure \ref{fig:spec-old}. The photometry is given in Extended Table~\ref{tab:phot}.
The wavelength units are in \micron\ (observed frame) and the flux units are in $10^{-19}$ ergs cm$^{-2}$ s$^{-1}$ \AA$^{-2}$ s$^{-1}$.

\bmhead{Code Availability}
All software packages used in this analysis are publicly available. In particular FAST++, Prospector-$\alpha$  and the {\tt hmf} python module are available on GitHub.

\bibliography{bibliography}

\clearpage

\captionsetup[table]{name=Extended Data Table}
\captionsetup[figure]{name=Extended Data Figure}

\section*{Methods}

\subsection*{NIRSPEC data reduction and calibration}

We observed the galaxy ZF-UDS-7329 with  JWST  on August 1$^{\hbox{st}}$ 2022 using the NIRSpec instrument 
in the MSA mode \citesupp{NIRSpec2022} as part of  our program \#2565 whose goal was
to secure spectroscopic redshifts of all objects in S18 which eluded deep ground-based spectroscopy. The PRISM disperser was used
producing spectra covering the wavelength range 0.6--5.0\micron\ at a spectral resolution ranging from 40 to 350 from blue to red ends. 
The galaxy was observed in three dither positions using a 5-shutter nod pattern (which has a nod distance large enough to avoid self-subtraction) and was close to centred transversely in the slit.
We reduced the data to 2D flux calibrated spectra using the STScI pipeline v1.12.5 with CRDS context {\tt jwst\_1152.pmap}
and used MSAEXP v0.6.18 to extract the 2D to 1D spectra.

Absolute spectrophotometric calibration of NIRSPEC data is difficult due to the narrow 0.2 arcsec slit and the wavelength-dependent point spread function (PSF) that causes significant 
additional aperture loss. Our calibration strategy was as follows: first we wished to verify that the spectrophotometry {\it through the slit\/} agreed with NIRCAM photometry of the same
galaxy. To do this we reduced the data in the pipeline's `point source calibration mode' but remove the {\tt pathloss} step that corrects for de-centering and point sources
and instead manually apply the {\it uniform} pathloss correction. Because of the software definitions of these quantities (\citesupp{Ferruit2016a}, p.10) the effect of this is to remove the point source aperture correction and 
return the flux though the slit. This we can then compare using synthetic slit photometry on NIRCAM images; because the PSF is dominated by the telescope it will be very similar in NIRCAM
and NIRSPEC. This comparison is showing in Extended Data Figure~\ref{fig:fluxcal} and it is excellent with a median offset of 0.13 mags and an RMS residual of $\sim 0.03$ mags. (We expect
small differences as the PSF would not be exactly the same and there will be small pointing errors). Given this validation of the 
slit calibration we next correct for aperture loss by fitting a second
order polynomial to the ratio of the total NIRCAM photometry to the slit photometry. In general we expect a wavelength dependence at this step as the slit photometry now includes
PSF aperture losses; however because the source is very uniform in colour and resolved this is only a small effect. We use the F150W$-$F277W colour which straddles the 4000\AA\ break to quantify this, correcting to total makes the colour 0.1 mags bluer which makes the spectral age younger, not older.

Finally we validated the formal propagation of uncertainties in the pipeline. We did this by comparing the individual spectra at the 3 dither positions and computing the RMS between them vs wavelength (after photometrically calibrating them to have the same overall shape). We expect on average this to approximately equal $\sqrt{2}\times$ the uncertainty, and we do indeed find this broad agreement; noting this
is a noisy comparison as there are a large number of outliers on single dither spectra due to defects or cosmic rays in individual exposures. We also find excellent
$\chi^2$ in the fitting when using the pipeline uncertainties which supports the noise model being correct.

ZF-COS-20115\cite{G17,S18} was also re-observed with JWST as a MSA filler in our program; the spectrum was reduced and calibrated using an identical procedure and serves as a useful
comparison for the  more typical $3<z<4$ post-starburst massive quiescent galaxy.

\subsection*{NIRCam analysis} \label{sec:NIRCAM}
NIRCAM images were taken as part of the PRIMER survey (program GO \#1837); we used the reductions and global photometry from the DAWN JWST archive\citesupp{DJA} v6 data release in a 0.7
arcsec diameter aperture. We provide the NIRCAM photometry in  Extanded Data Table~\ref{tab:phot}.

To measure the object size we  perform two-dimensional fits to the surface brightness distributions of the NIRCam F444W galaxy images using a Python package, {\tt Galight} \citesupp{galight}. With {\tt Galight}, we prepare a cutout galaxy image, detect neighboring sources in the field of view; {\tt Galight} allows for simultaneously fitting of as many neighboring objects as needed, in order to avoid source blending. 
We fit a single-component S\'{e}rsic profiles to the two dimensional surface brightness distributions of all detected objects, making use of galaxy imaging data, the noise level maps, and custom-made PSF models. The fit parameters are total magnitudes ($M$), half-light radius ($r_{e}$), S\'{e}rsic index ($n$), and axis ratio ($b/a$). With the input ingredients, {\tt Galight} convolves the theoretical models (i.e., S\'{e}rsic profile) with the Point Spread Function (PSF) before fitting them to the galaxy images. We define the effective radius as the semi-major axis of the ellipse that contains half of the total flux of the best-fitting S\'{e}rsic model. We find the effective radius to be $r_{e}=1.15^{+0.08}_{-0.08}~\mathrm{kpc}$, the Sersic index to be $n=2.41^{+0.24}_{-0.29}$ and the axis ratio
$(b/a) =0.33^{+0.01}_{-0.01}$ --- all consistent with the visual appearance of an edge on disk which is shown in Figure~\ref{fig:spec}.

\subsection*{Spectroscopic Line Spread Function}

The NIRSpec PRISM disperser has a very low and non-linear spectral resolution and dispersion scale. This coupled with the high SNR per pixel
of the spectrum  makes it quite important to account for systematic errors in the fitting. In particular the 
wavelength-dependent line spread function (LSF) is significant and without accounting for this we can not obtain
acceptable fits, especially around the 4000\AA\ break. 
To calculate the LSF we first need a model for the image size, as the galaxy is smaller than the slit width. This we obtain from the PRIMER images
(which has very closely the same PSF as the NIRSpec focal plane), 
we supersample them and project them along the slit to obtain a 1D profile across the slit in each band. We find these projected profiles are well approximated by
a gaussian function with  a small and smooth increase in effective image size from 0.16 to 0.18 arcsec (Full Width Half Maximum) with wavelength due to the 
telescope's diffraction. This increase  is well approximated by a linear function. Multiplying this by the PRISM dispersion 
function\citesupp{PRISM-disp} gives the LSF width as
a function of wavelength.

\subsection*{Spectroscopic redshift determination}

We use {\tt slinefit} (S18) which fits stellar and emission line templates, We use 
version 2.3 where we added the feature to include a wavelength dependent LSF  in the modelling. 
We obtain a redshift of $z=3.205 \pm 0.005$ and a template fit very similar to those we found later with full stellar population modelling. We derived this redshift error
using Monte Carlo simulations taking
a 1.5 Gyr old Bruzual \& Charlot 2003 `Simple Stellar Population' model\cite{BC03}, which approximates the spectrum of our galaxy, varying the redshift from 3.19 to 3.23 and creating simulated NIRSPEC spectra including LSF and pixelisation. We find
the typical redshift recovery with {\tt slinefit} is  $\pm 0.005$ over this range; which is about $\pm 1/8$th of a spectral pixel at 2\micron\ which seems reasonable for high SNR full spectrum fitting.
The accuracy gets considerably worse if the LSF is ignored. We also estimate the effect of the redshift error on the spectral modelling which is a potential 
concern given our high SNR but coarse resolution. We do this by shifting and re-simulating the model; we find a redshift error of 0.01  introduces systematic flux residuals at the 1\% level which
is below that of our other noise sources.

\subsection*{FAST++ and Prospector fitting}

We use FAST++ version 1.5.0 which was enhanced by us to include  a wavelength dependent LSF in the modelling. The spectral libraries used by FAST++ are only well characterised at spectroscopic resolution in the rest-frame optical; therefore we limit the fitting to 
rest frame wavelengths $<0.8$\micron. We use the same  general star formation history (SFH) models as S18 and identical
parameter ranges (as listed S18 Table 1);  modelling the SFH as an exponential rise and fall at earlier times 
plus a component of recent residual star formation. The latter is 
often useful when there is a superposition of old and young stellar populations which can bias mass and age estimates; however
we find the young component is negligible in the case of ZF-UDS-7329. We allow metallicity to vary up to $2.5\times$ solar  and 
adopt a Chabrier Initial Mass Function\citesupp{Chabrier2003}. The underlying models utilise standard solar-scaled element abundances. Burst
times are allowed to go as short as 10 Myr which can be important for unbiassed estimates of quenching rates\cite{Seuss2022}. For each model
evaluation we compute the time at which 50\% of the stars had formed for the trial set of parameters.
We marginalise over the model grid and determine errors on model parameters, derived parameters and on the SFH at that time step using 1000 Monte-Carlo simulations. 

For Prospector we use the same LSF model as for FAST++. We corrected an issue in the interpolation of the high-resolution
spectral models which caused incorrect pixel downsampling to our low spectral resolution.\citesupp{prospector-problem} 
Our version is provided in a GitHub fork \citesupp{prospector-fix} . 
We limit fitting to  rest frame wavelengths $<0.7\micron$;  as recommended for these stellar libraries according to the Prospector documenatation. 
We use a {\tt continuity\_sfh} to model the formation history of our galaxy\citesupp{leja2019a} with 7 custom time bins. These follow the default
approximately logarithmic sampling at young ages, important for capturing potential recent star formation, but we impose finer sampling of
0.25 Gyr at ages $>$ 1 Gyr to capture the epoch of maximum star formation more carefully.
We additionally let the stellar metallicity ($-2 \leq \log_{10}(Z/Z_\odot)  \leq 0.19$) and the dust optical depth ($0 \leq \tau \leq 2.0$) vary as free parameters.
Our bins are fixed in time; simulations have shown that this choice\cite{Seuss2022}  can accurately recover ages but have difficulty recovering 
the duration of recent quenching episodes due to rounding errors at bin edges; however our quenching proves not to be recent and we are
not trying to accurately resolve in time the quenching event. Higher spectral resolution data would allow more flexible models to be explored.
As shown in the main text given our choices we find strong agreement between parametric and non-parametric approaches.

Since we do  explicit calibration of the data to the photometry, we do not consider the similar options in FAST++ and Prospector 
that allow an extra polynomial function to be fitted to do this. In both cases we include the 1--5\micron\ PRIMER  photometry in the fits though this makes
little practical difference as the weights are dominated by very high SNR spectrum.

For both codes we we find reduced $\chi^2\sim 2$ in the best fits which is excellent given there are $\sim 100$ independent spectral pixels each with SNR of 50--70. (The models
have 7--9 free parameters.)
The $\chi^2$ gets much worse if we do not include the LSF in the fit or do the downsampling incorrectly.

Extended Data Figure~\ref{fig:sfh_Z05} shows the effect of force fitting the [Fe/H]$=$0.05 solution which is formally excluded. As expected the age can come down to $\sim 1 $ Gyr but a
substantial amount of star formation is still preferred at ages $>1.3$ Gyr and the age distribution of star formation remains significantly older than in the TNG300 galaxies.
There is more disagreement between FAST++ and Prospector which we attribute to the effect of the mismatch of the supersolar model to the spectrum.

\subsection*{Rest frame near-infrared spectrum}

Our model fitting excludes rest-frame wavelengths $>0.8$\micron\  as the stellar libraries that underpin the models are currently poorly determined in this region. We show this
wavelength range in Extended Data Figure~\ref{fig:spec-old}. The extrapolation of the best fitting Prospector model is overlaid on the figure; though it reproduces the overall spectral shape the detailed spectral variations are not reproduced. 

However we can still see if the spectrum in this wavelength range is qualitatively consistent with an old population by comparing empirically with nearby galaxy templates. We
do indeed notice several features consistent with the presence of older stellar populations. Most noticeable is the broad absorption bump at 0.94 \micron. This feature is known, but not well represented in stellar libraries, as at $z=0$ the wavelength range
overlaps with significant telluric absorption \citesupp{Riffel+2015}. It is thought to be due 
 to a mixture of ZrO, CN and TiO features in Thermally Pulsing Asymptoptic Giant Branch stars \citesupp{Riffel+2015}.
 Empirically this bump appears in stellar populations older than a Gyr. 
 
We  compare with low redshift galaxies from \citesupp{Mason+2015}. This sample
is near-infrared spectra 
 of nearby galaxy nuclei many of which harbour weak Active Galactic Nuclei (AGN) so this is not ideal. We choose two example objects from here that 
are classified as dominated by old stellar populations and with no AGN emission lines. 
 NGC 205 is an intermediate age ($< 1$ Gyr) post-starburst; it has a hotter
 spectrum, and a weaker bump than ZF-UDS-7329. In contrast the spectrum of NGC 5850 (with age $\sim$ 5 Gyr) has a much stronger bump qualitatively similar
 to what we see. The CN feature at rest frame 1.1\micron\ demonstrates similar behaviour. Overall the rest-frame near-infrared spectrum of our galaxy is consistent with ages
 older than a Gyr.
 
 The current lack of coverage of the near-infrared templates is due to the issues of spectroscopy from the ground at wavelengths with significant atmospheric absorption. This sitation
 is likely to improve rapidly due to the capabilities of JWST.

\subsection*{Halo masses}
 
 The halo mass functions were computed using the python {\tt hmf} module v3.4.4 which implements a wide variety of analytic predictions. We found that at
 our high redshifts the resulting mass functions did not vary significantly with choice of model and agreed well with the results from the large
 volume FLAMINGO simulation.
 
 We adopt a cosmology of $\Omega_m = 0.3$,  $\Omega_\Lambda = 0.7$, $H_0 = 70$ km$^{-1}$ s$^{-1}$ Mpc$^{-1}$ for time and halo mass analytic calculations. We note
 that there is a small difference in FLAMINGO; with the main one being the use of $H_0 = 67$ km$^{-1}$ s$^{-1}$ Mpc$^{-1}$ but this has negligible impact on the comparison.


\setcounter{figure}{0}    

\clearpage

\begin{table}
\tabcolsep = 1.8cm
\begin{tabular}{lrr}
\toprule
Band & mag \\
\midrule
F115W & $25.567 \pm 0.034$ \\
F150W & $24.299 \pm 0.009$ \\
F200W & $22.826 \pm 0.002$ \\
F277W & $22.154 \pm 0.001$ \\
F356W & $21.827 \pm 0.001$ \\
F410M & $21.616 \pm 0.001$ \\
F444W & $21.546 \pm 0.001$ \\
\bottomrule
\end{tabular}
\caption{NIRCAM integrated photometry for ZF-UDS-7329. Note the tabulated random errors are small for such a bright galaxy; it is estimated that systematic
errors are of order 0.02 mags.\protect\citesupp{JWSTcal}} \label{tab:phot}
\end{table}

\begin{figure}%
\includegraphics[width=14.5cm]{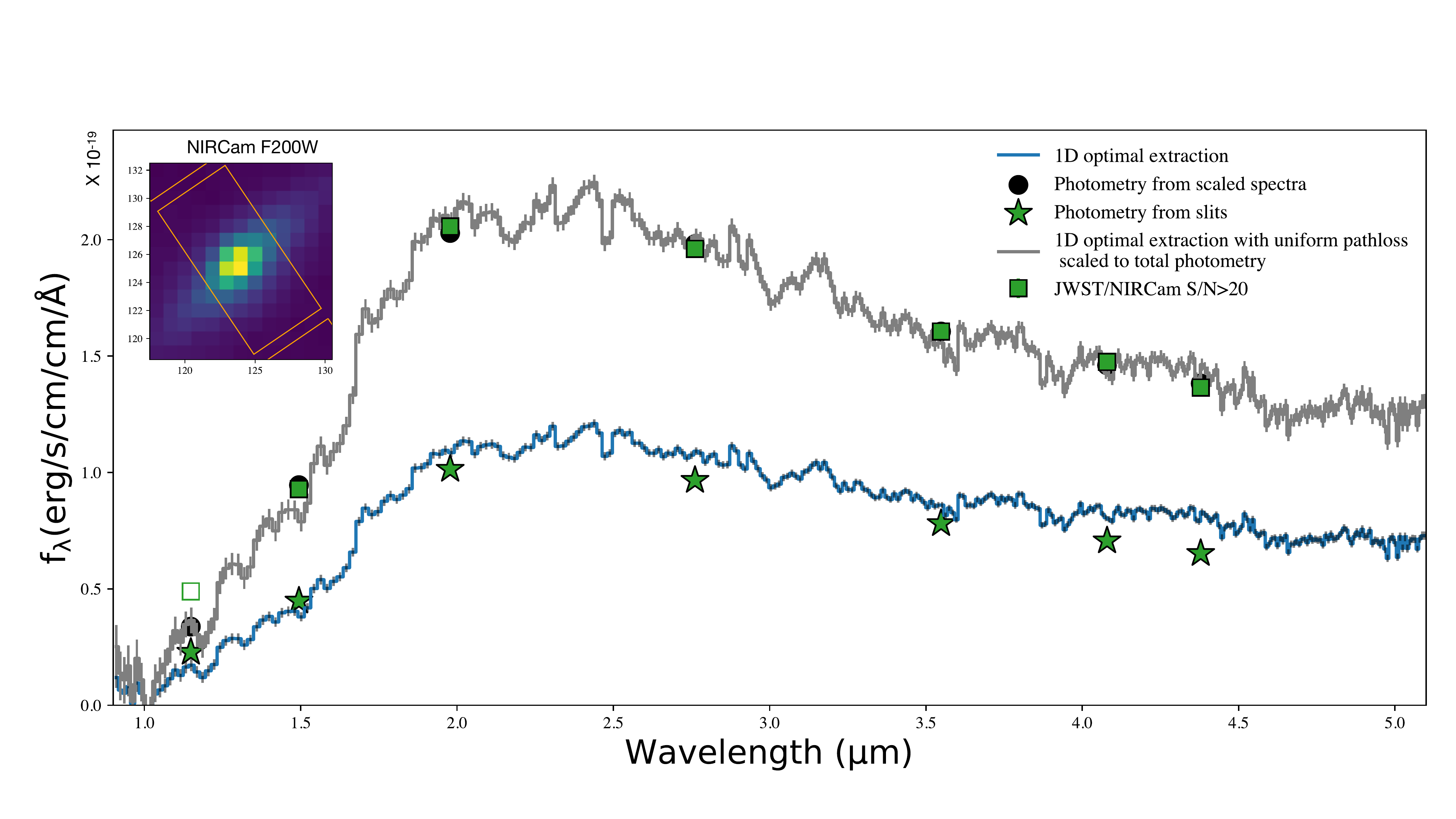}
\caption{Flux calibration of the NIRSpec spectrum. The lower blue curve shows the derived flux through the slit, compared with NIRCAM photometry with a synthetic slit aperture. This
shows very good agreement. The upper grey curve shows the correction of the spectrum to total NIRCAM photometry. Both spectra show have flux error bars superimposed. The inset shows the MSA shutter footprint overlaid on the NIRCAM
F200W image. }
\label{fig:fluxcal}
\end{figure}

\begin{figure}%
\includegraphics[width=14.5cm]{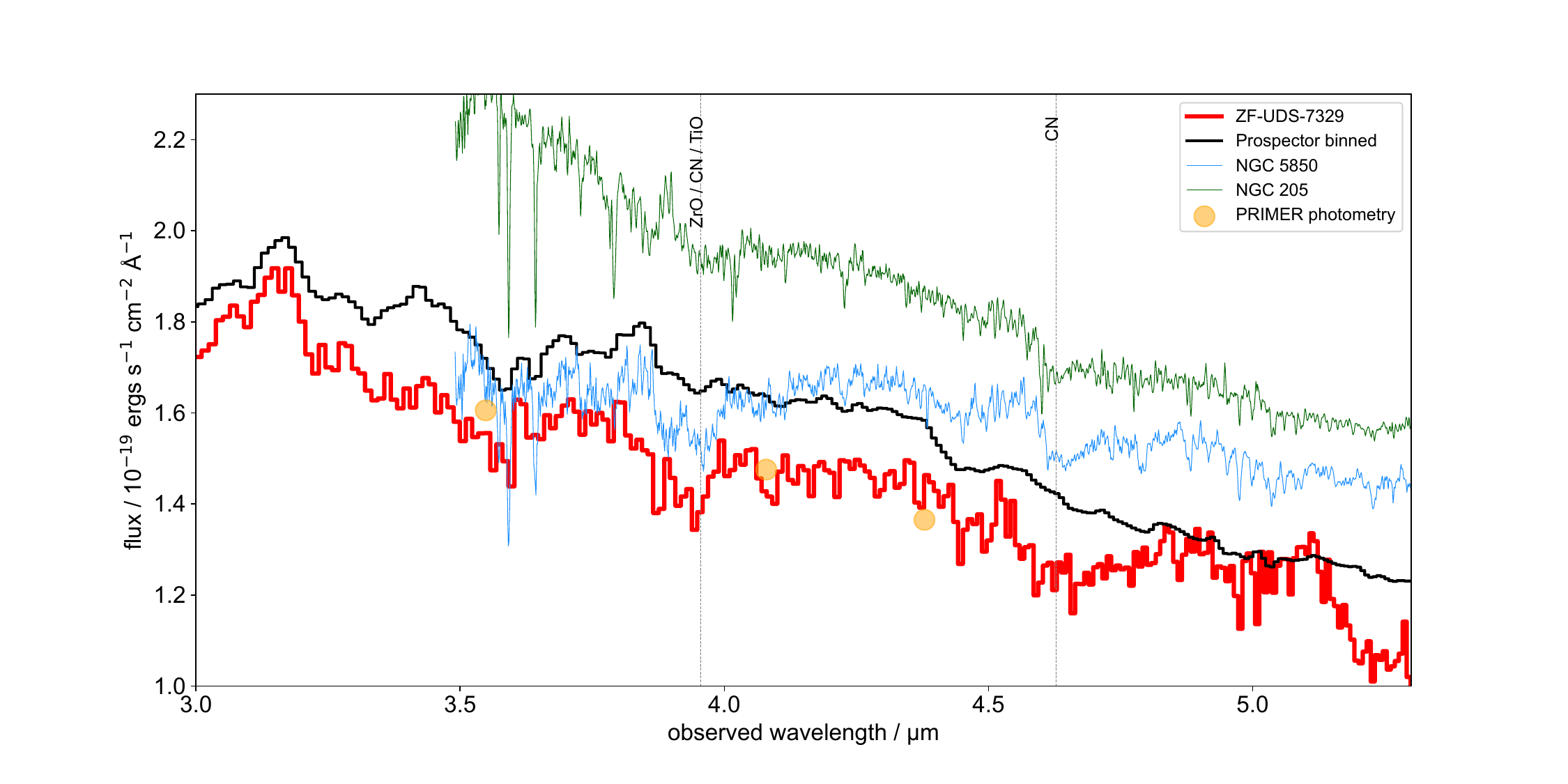}
\caption{Comparison of spectra at longer wavelengths.
ZF-UDS-7329 in the rest frame 0.7--1.3\micron\ region is compared to redshifted high signal:noise spectra\protect\cite{Mason+2015} 
of the nuclei of local NGC galaxies (these are normalised to the same flux at 0.9--0.95\micron\ rest and then offset for clarity). NGC 5850 is a nearby spiral with a 
luminosity weighted age of $\sim 5$ Gyr, and it can be seen that the
0.94\micron\ absorption (ZrO, CN, TiO bands)
 is quite similar, and there is overall a very good match between them to the bumps and wiggles in the continuum which arise 
from numerous molecular bands in cool stars. NGC205's light is dominated
by an intermediate age population (0.1--1 Gyr) and it can be seen that the 0.94\micron\ feature, 
and other molecular bands are much weaker. Note the flux axis is greatly zoomed compared to Figure~\ref{fig:spec} to highlight very weak
absorption features. 
\label{fig:spec-old}}
\end{figure}

\begin{figure}%
\includegraphics[width=14.5cm]{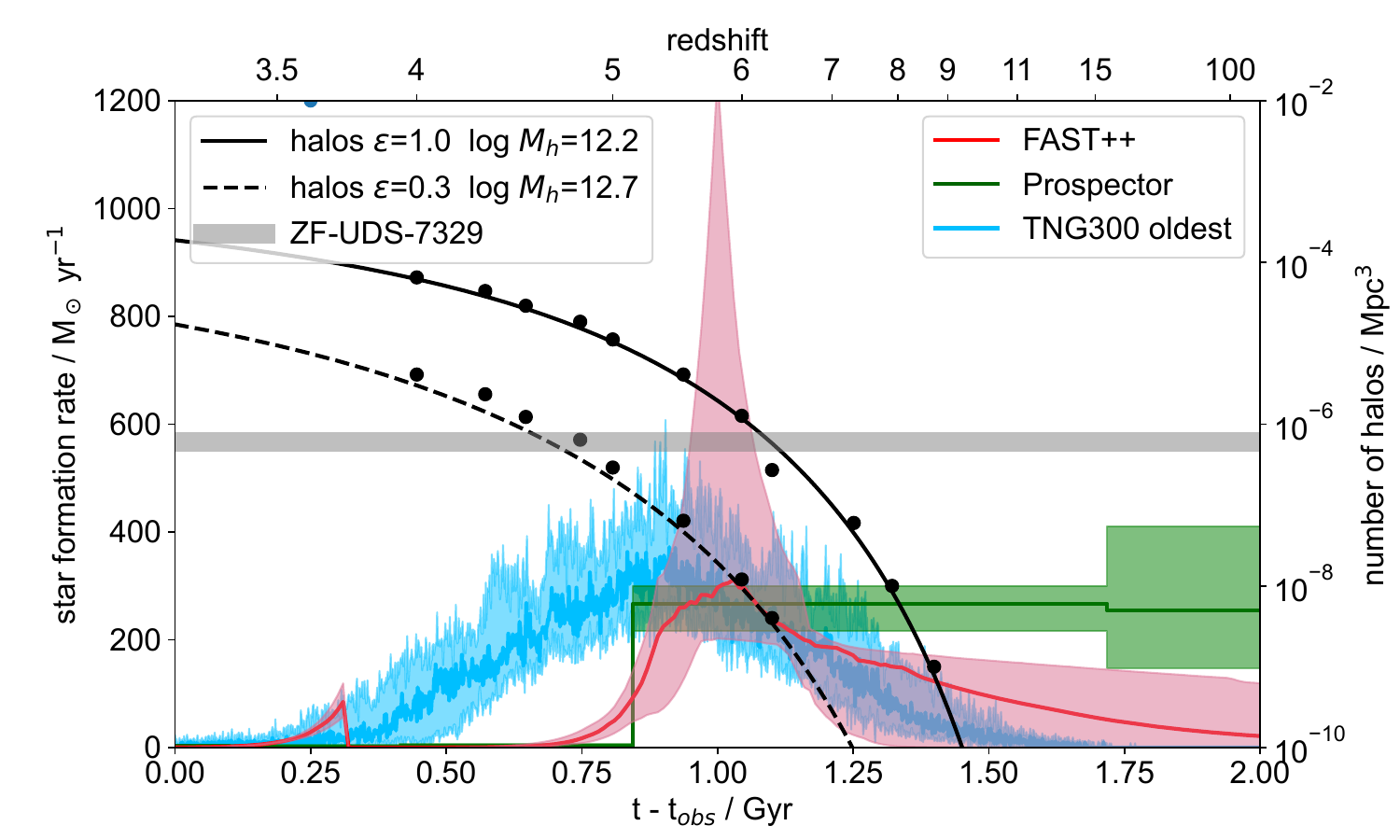}
\caption{Comparison of star formation histories with dark matter halo assembly histories for high metal abundance.
This is a version of Figure~3 where we force fit the poor fit model with [Fe/H] = 0.05. The age is reduced to $\sim 1$ Gyr but is still significantly discrepant
with theoretical expectations. Colours, lines and symbols are as in Figure~3.
\label{fig:sfh_Z05}}
\end{figure}

\clearpage

\bibliographystylesupp{sn-mathphys}
\bibliographysupp{bibliography}

\section*{Author Information}

The authors declare that they have no competing financial or non-financial interests. Correspondence and requests for materials should be addressed to Karl Glazebrook (\href{mailto:kglazebrook@swin.edu.au}{kglazebrook@swin.edu.au}). \\

\noindent Reprints and permissions information is available at \href{https://www.nature.com/reprints}{www.nature.com/reprints}.

\end{document}